\documentclass{article}  
\usepackage{breckenr2005}

\newcommand{\pt} {\mbox{$p_{T}$}}
\frompage{000} \topage{000}                                              
\def\rightmark{Identified Particle Production at RHIC}
\def\leftmark{Felix Matathias for the PHENIX collaboration}
 
\title{Identified Particle Production in p+p, d+Au, and Au+Au Collisions at RHIC} 
\authors{
{Felix Matathias$^1$ for the PHENIX Collaboration %
}\\[2.812mm]
{\normalsize
\hspace*{-8pt}$^1$ Columbia University, Nevis Laboratories, \\
PO Box 137, Irvington, NY 10533, U.S.A.\\[0.2ex] 
}}
 
\abstract{Measurements of identified particle production with the PHENIX experiment at RHIC 
have reached a mature state, where a multitude of nuclear systems at different colliding energies 
have been studied. The discovery configurations of $\sqrt{s_{NN}}$ = 130 and 200 GeV Au+Au collisions 
have now been supplemented by additional Au+Au and Cu+Cu configurations at various energies, along with 
baseline p+p and d+Au runs at $\sqrt{s_{NN}}$ = 200 GeV. In this work we present a systematic study 
of the Cronin effect in d+Au collisions and recent results from p+p collisions.
We then proceed to make a critical comparison of pion, kaon and proton production in heavy ion
and baseline systems, and discuss the observed nuclear effects on hadron production.  }

 
\begin{document}
 
\maketitle
\setcounter{page}{1}

\section{Introduction}\label{intro}

The abundance and composition of particles that reached the PHENIX detectors after central collisions of Au nuclei
at $\sqrt{s_{NN}}$ = 130 and 200 GeV generated a lot of excitement in the Relativistic Heavy Ion community 
soon after the Relativistic Heavy Ion Collider (RHIC) assumed operations. The transverse momentum spectra of pions 
were greatly suppressed by factors of 4-5 at high \pt~\cite{ppg003,ppg014} compared to expectations from point-like scaling, and the proton
and antiproton \pt\ spectra, in contrast to any expectation or prediction, crossed the pionic spectra at 
around 2 GeV in \pt~\cite{ppg006,ppg015,ppg026}. For the first time, matter created in the laboratory had reached such exotic conditions 
of temperature and pressure that the creation of a pion or a proton were equally probable, despite their mass difference
and quark content. Suppression of high energetic particles had been long expected to be the smoking gun signature of the creation 
of Quark Gluon Plasma~\cite{Gyu90,BDMPS}, where hard-scattered partons suffer energy loss as they propagate through the hot and dense medium. 
The particle composition anomaly though was a complete surprise and
to this day constitutes one of the great discoveries of the relativistic heavy ion programme at PHENIX, and 
RHIC in general.

Many theoretical works assumed the task of explaining these remarkable properties of super-heated partonic matter
and we now have a much firmer theoretical understanding of the underlying dynamics of heavy ion collisions at
RHIC energies, but due to the hugely successful performance of the RHIC collider in the next 4 years after the first discoveries,
the wealth of experimental data are still at the forefront of the community's efforts to solve the Quark Gluon Plasma puzzle.
With the successful completion of the fifth year of RHIC operations in 2005, which provided Cu+Cu collisions and polarized p+p collisions,
PHENIX has recorded data on Au+Au collisions at  $\sqrt{s_{NN}}$ = 19.6, 62.4, 130, and 200 GeV, 
Cu+Cu collisions at $\sqrt{s_{NN}}$ = 22.5, 62.4, and 200 GeV, along with baseline measurements in
p+p and d+Au collisions at  $\sqrt{s_{NN}}$ = 200 GeV. 

The novel effects observed in central Au+Au collisions
require a detailed account of the initial state conditions. 
The effects from the initial state are best studied by performing a control experiment 
in which only cold nuclear matter is produced. Deuteron + gold collisions 
at $\sqrt{s_{NN}}$= 200 GeV serve this purpose. Since hot and dense final state 
medium is not created in these type of collisions, the initial state conditions become accessible 
to the experiment. Known initial state effects include nuclear shadowing 
and gluon saturation~\cite{colorglas}, along with the Cronin effect~\cite{cronin,straub,antreasyan}, which enhances the hadron 
yields, and, therefore, acts in the opposite direction compared to the suppression of particle yields in central Au+Au collisions.
The Cronin effect is usually attributed to momentum broadening due to multiple
initial state scattering~\cite{LevPetersson,Accardi,PappLevai,Vitev,Wang,boris} but the species dependence of the effect is not yet
completely understood and further experimental study is interesting in its own right. 
Since it has been observed that at lower energies, Cronin enhancement is stronger for 
protons than for pions~\cite{antreasyan}, this effect has to be considered at RHIC energies
before new physics for baryon and meson production is invoked.

Recently, Hwa and collaborators provided an alternative explanation due to final state interactions. 
The particle species dependent enhancement is attributed to recombination 
of shower quarks with those from the medium, where no distinction is made if hot or 
cold nuclear  matter is produced~\cite{Hwa}.
Identified hadron production measured as a function of centrality brings important experimental 
handles to this long outstanding problem. The dependence of the enhancement upon 
the thickness of the medium can differentiate among the scattering models, and 
the species dependence helps to separate initial from final state effects in d+Au.

\section{Experiment and Data Analysis}\label{techno}  

Data presented here include collisions at $\sqrt{s_{NN}} = 200$~GeV 
of Au+Au taken in the 2002 run of RHIC, and d+Au and p+p collected
in 2003. In the following we discuss analysis of the p+p and d+Au data;
details of the Au+Au analysis are found in~\cite{ppg026}. 
Events with vertex position along the beam axis within  $|z| < $30 cm
were triggered by the Beam-Beam  Counters (BBC) located 
at $|\eta|$~=~3.0-3.9~\cite{nim_phenix}. The minimum bias trigger 
accepts 88.5 $\pm$ 4\% of all d+Au collisions that satisfy the vertex condition, 
and 51.6 $\pm$ 9.8\% of p+p collisions. A total of 42 $\times$ 10$^6$ d+Au events 
and 25 $\times$ 10$^6$ minimum bias p+p events were analyzed.

In p+p collisions, differential invariant cross section is calculated by Eq.~\ref{eq:goldenformula}
\begin{equation}
E\frac{ d^{3}\sigma } { dp^{3} } =  \frac{\sigma_{BBC}}{N^{Total}_{BBC}}  \cdot \frac{1}{2\pi} \cdot \frac{1}{p_T} 
                           \cdot C^{geo}_{eff}(\pt)  \cdot \frac{1}{C^{BBC}_{bias}} \cdot  \frac{d^2N}{ dp_{T} dy}
\label{eq:goldenformula}
\end{equation}

The BBC trigger cross section, $\sigma_{BBC}$, was determined via the van
der Meer scan technique and it was measured to be 
$\sigma_{BBC} =  23 \, \pm \, 2.1(9.6\%) \,\, mb$.
The factor $C^{geo}_{eff}(\pt)$ denotes the efficiency and geometrical acceptance correction, 
calculated with a detailed GEANT Monte Carlo simulation of the PHENIX aperture.
$C^{geo}_{eff}(\pt)$ normalizes the cross section in one unit of rapidity and full azimuthal coverage.
The $C^{BBC}_{bias}$ factor corrects for the fact that the forward BBC trigger counters
see only a fraction of the inelastic p+p cross section. This subset of 
events that the BBC triggers on, contains only a fraction of the inclusive
particle yield at mid-rapidity. For charged hadrons this factor was determined 
to be 0.80 $\pm$ 0.02, independent of \pt, using the data from the beam bunch crossing triggers.

There were no van der Meer scans performed during the 
d+Au run, and therefore PHENIX only measures the inelastic yield per BBC triggered event. 
The collision centrality is selected in d+Au using the south 
(Au-going side) BBC (BBCS). We assume that the BBCS signal is 
proportional to the number of participating nucleons ($N^{Au}_{part}$) 
in the Au nucleus, and that the hits in the BBCS are uncorrelated 
to each other. Using a Glauber model~\cite{glauber} and simulation
of the BBC, we define 4 centrality classes in d+Au collisions, as 
was discussed in detail in~\cite{ppg036}. The mean number of binary
collisions that correspond to each centrality bin are 15.4$\pm$1.0, 
10.6$\pm$0.7,  7.0$\pm$0.6 and  3.1$\pm$0.3 for the most peripheral d+Au bin.
For the Minimum Bias d+Au collisions $\langle N_{coll} \rangle$ = 8.5 $\pm$ 0.4.  

Charged particles are reconstructed using a drift chamber (DC) and 
two layers of multi-wire proportional chambers with pad readout
(PC1, PC3)~\cite{nim_phenix}. Particle identification is based on particle mass calculated from 
the measured momentum and the velocity obtained from the 
time-of-flight and path length along the trajectory. The measurement uses the 
portion of the east arm spectrometer containing the high resolution
time-of-flight(TOF) detector, which covers pseudo-rapidity 
 $|\eta| = 0.35$ and $\Delta\phi=\pi/8$ in azimuthal angle.    
The timing uses the BBC for the global start, and stop signals
from the TOF scintillators located at a radial distance of 5.06 m.
The system resolution is $\sigma \approx$ 130 ps.

Corrections to the charged particle spectrum for geometrical 
acceptance, decays in flight, reconstruction
efficiency, energy loss in detector material,
and momentum resolution are determined using a 
single-particle GEANT Monte Carlo simulation. 
The proton and antiproton spectra are corrected for feed-down
from weak decays using a Monte Carlo simulation which uses 
experimental data as input for particle composition and the $\Lambda$ 
spectrum from~\cite{UA5}, \cite{heinzpaper}, \cite{heinzppt}, \cite{cai}, \cite{mtscaling}.
The Monte Carlo is used to decay and propagate the products 
of the weak decays through the PHENIX magnetic field and central arm detectors. 
The resulting feed-down proton and antiproton spectra are then subtracted from the 
inclusive measured spectra. The contribution from feed-down protons is approximately 
30$\%$ at moderate and high \pt\, increasing slowly to
40$\%$ at $p_T$ = 0.6 GeV/c. 

\subsection {Systematic Uncertainties}

Systematic uncertainties on the hadron spectra arise from the small
remaining contamination by other species, backgrounds remaining
after the matching hit requirement in the TOF, and residual 
time variations in the TOF timing.  Uncertainties
due to particle identification cuts are momentum dependent. For
protons and antiprotons, the identification uncertainty is 8\% 
at low $p_T$ and decreases to 3\% at high $p_T$. Kaons at low
momentum have 10\% PID uncertainty, decreasing to 3\% at high $p_T$.
For pions the uncertainty increases from 4 to 10 \% with increasing
$p_T$. The systematic error on the feed-down
proton spectrum is 24$\%$, primarily due to uncertainty in the 
measured $\Lambda$ spectra and particle composition.
The resulting systematic error on the final prompt proton and antiproton
spectra is of the order of 10$\%$ in both p+p and d+Au.
The systematic error on the proton to pion ratio is 12$\%$, 
including the uncertainty on $\overline\Lambda/\Lambda$.

Systematic uncertainties on the d+Au nuclear modification factors mostly cancel
as the p+p and d+Au data were collected immediately following one another,
and detector performance was very similar.
The overall systematic error in the nuclear modification factor is 
due to uncertainties in the reconstruction efficiencies, fiducial
volumes, and small run-by-run variations. It is approximately
10\%, independent of \pt. An additional d+Au scale uncertainty, 
which is shown as boxes in the following figures, 
is the quadrature sum of uncertainties on the p+p cross section of 9.6\%, 
and the number of binary collisions in each centrality bin.

The systematic error on the Au+Au nuclear modification factors is
derived by propagating the systematic errors
on p+p and Au+Au data~\cite{ppg026} to the final ratio. The
average systematic error for pions is approximately 15\%, while
for protons and antiprotons it is on the order of  19\%.
The normalization uncertainty, as in d+Au, is the quadrature sum of 
uncertainties on the p+p cross section and the number of binary collisions in
the corresponding Au+Au centrality bin. 
For the most central Au+Au bin, $N_{coll}$=1065.4$\pm$105.3, 
while for the most peripheral centrality bin,  $N_{coll}$=14.5 $\pm$ 4.0 .

\section{Results}\label{results}

\begin{figure}[htb]
\vspace*{-.2cm}
                 \insertplot{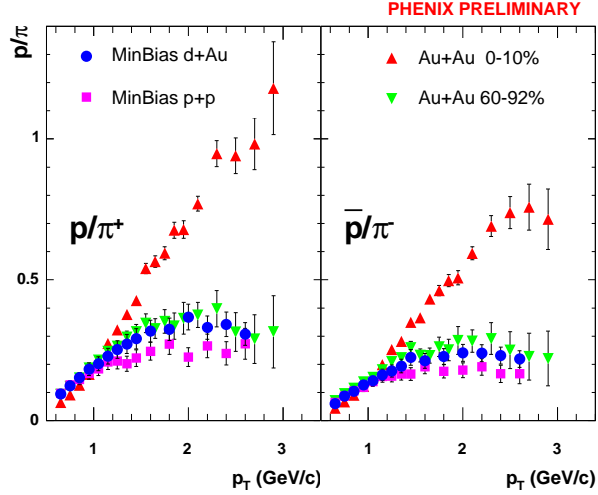}
\vspace*{-1cm}
\caption[]{The ratio of protons to $\pi^+$ and antiprotons to
$\pi^-$ in minimum bias p+p and d+Au compared to peripheral
and Au+Au collisions. Statistical error bars are shown.}
\label{fig:ptopi} 
\end{figure} 

\subsection{Proton to Pion Ratio}
The proton to pion ratio from minimum bias p+p and minimum bias d+Au 
are compared to each other and to central and peripheral Au+Au 
collisions in Figure~\ref{fig:ptopi}. As noted above,
protons and antiprotons are feed-down corrected in each system.

The $p/\pi$ ratio in d+Au is very similar to that in peripheral Au+Au 
collisions, and lies slightly above
the p+p ratio. The $p/\pi$ ratio in central 
Au+Au collisions is, however,  much larger. The difference between the 
ratio in d+Au and central Au+Au clearly indicates that baryon yield
enhancement is not simply an effect of sampling a large nucleus in the
initial state. The large enhancement requires the presence of a substantial 
volume of nuclear medium with high energy density.

\subsection {Nuclear Modification Factors}

\begin{figure}[htb]
\vspace*{-.2cm}
                 \insertplot{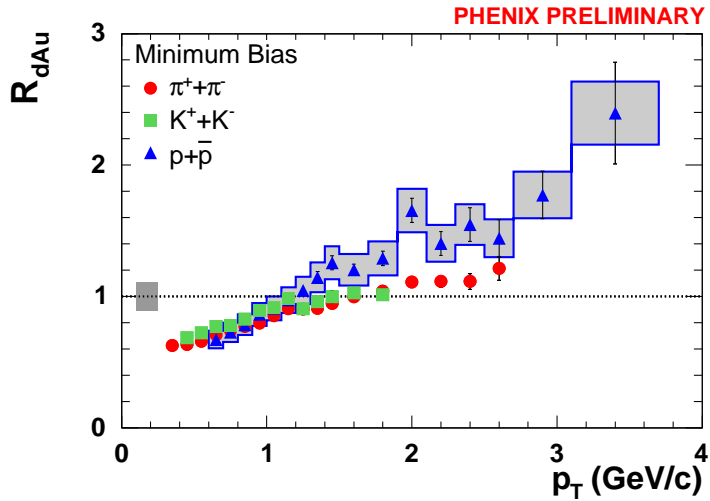}
\vspace*{-1.cm}
\caption[]{ Nuclear modification factor $R_{dA}$  for pions, kaons 
and protons in d+Au collisions for Minimum Bias events.
The error bars represent the statistical errors, while the
systematic errors, which are dominated by uncertainties in the 
absolute yield and calculation of $N_{coll}$, are shown as
vertical bars. Point to point errors are also shown for the 
protons and antiprotons.}
\label{fig:rdauminbias}
\end{figure} 

The measurement of identified hadrons in both d+Au and p+p
collisions allows study of the centrality dependence of
the nuclear modification factor in d+Au.
A standard way to quantify nuclear medium effects on high $p_T$ 
particle production in nucleus-nucleus collisions is provided by the 
{\it nuclear modification factor}. This is the ratio of the 
d+A invariant yields to the scaled p+p invariant yields:
\begin{equation} 
R_{dA}(p_T)\,=\,\frac{(1/N^{evt}_{dA})\,d^2N_{dA}/dy dp_T}{T_{dAu}\, 
d^2\sigma^{pp}_{inel}/dy dp_T},
\label{eq:R_dA}
\end{equation}
where $T_{dAu} = \langle N_{coll}\rangle/\sigma^{pp}_{inel}$ describes the
nuclear geometry, and $d^2\sigma^{pp}_{inel}/dy dp_T$ for p+p collisions
is derived from the measured p+p cross section. 
$\langle N_{coll}\rangle$ is the average number of 
inelastic NN collisions determined from the Glauber 
simulation described above.

Figure~\ref{fig:rdauminbias} shows $R_{dA}$ for pions, kaons and
protons for the minimum bias d+Au centrality bin. We observe a 
nuclear enhancement in the production of hadrons with $p_T \ge$ 
1.5 - 2 GeV/c in d+Au collisions, compared to that in p+p. 
As was already suggested when comparing the enhancement for
inclusive charged hadrons with that of neutral pions~\cite{ppg028},
there is a species dependence in the Cronin effect. The
Cronin effect for charged pions is small, but non-zero, as
was observed for neutral pions. The nuclear enhancement for
protons and antiprotons is considerably larger, in agreement with~\cite{starpiddau}. 
The kaon measurement has a more limited kinematic range, but the
$R_{dA}$ is in agreement with that of
the pions at comparable $p_T$.

\begin{figure}[htb]
\vspace*{-.2cm}
                 \insertplot{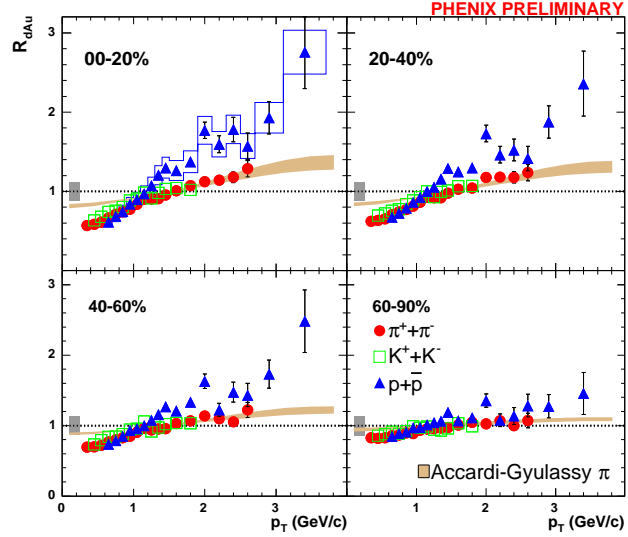}
\vspace*{-1cm}
\caption[]{ Nuclear modification factor $R_{dA}$  for pions, kaons 
and protons in d+Au collisions in four centrality bins.
The error bars represent the statistical errors, while the
systematic errors, which are dominated by uncertainties in the 
absolute yield and calculation of $N_{coll}$, are shown as
vertical bars. $p_T$ dependence of the systematic uncertainty
is minimal and included in the magnitude of the bar.
The solid bands show the calculation of the
nuclear modification factors for pions by Accardi and Gyulassy
~\cite{Accardi}.}
\label{fig:rdau}
\end{figure} 

Figure~\ref{fig:rdau} shows $R_{dA}$ for pions, kaons and
protons in the four d+Au centrality bins.
Peripheral d+Au collisions, which include an average of 3
nucleon-nucleon collisions, do not show modification of
high momentum hadron production, compared to that in p+p collisions.
At $p_T \le $ 1 GeV/c, the nuclear modification factor falls below
1.0. This is to be expected as soft particle production scales 
with the number of participating nucleons, not with the number 
of binary nucleon-nucleon collisions. More central collisions 
show increasing nuclear enhancement in both high $p_T$ pion 
and proton production.

The bands in Figure~\ref{fig:rdau} 
show a calculation of the Cronin effect for pions by Accardi
and Gyulassy, using a p-QCD model of multiple semi-hard collisions
and taking geometrical shadowing into account~\cite{Accardi}. 
The agreement above 1 GeV/c, where the calculation should be
reliable, is very good for all four centrality bins. 
The agreement illustrates the effects of multiple partonic
scattering and nuclear shadowing. The quantitative agreement
leaves very little room for dynamical shadowing effects in the 
nuclear initial state at mid-rapidity at RHIC.

\begin{figure}[htb]
\vspace*{-.2cm}
                 \insertplot{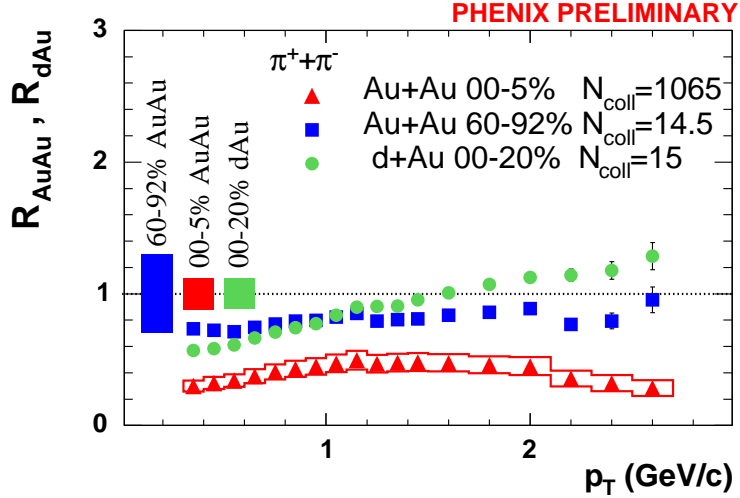}
\vspace*{-1cm}
\caption[]{ Nuclear modification factors for charged pions,
comparing central and peripheral Au+Au collisions to central d+Au.
It should be noted that the number of binary nucleon-nucleon 
collisions in peripheral Au+Au and central d+Au is very similar.
Solid bars on the left indicate $p_T$-independent normalization 
uncertainties on the nuclear modification factors for the three
systems; error bars indicate statistical errors only.
}
\label{fig:rauaudaupi} 
\end{figure} 

\begin{figure}[htb]
\vspace*{-.2cm}
                 \insertplot{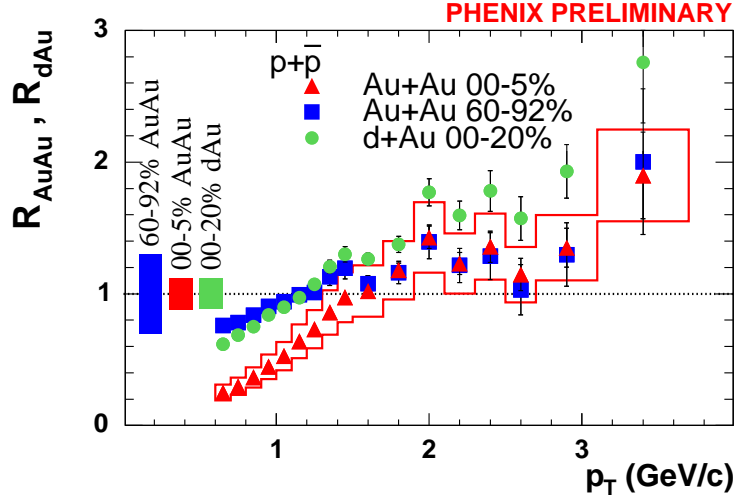}
\vspace*{-1cm}
\caption[]{ Nuclear modification factors for protons and antiprotons,
comparing central and peripheral Au+Au collisions to central d+Au.
}
\label{fig:rauaudaup} 
\end{figure} 


\begin{figure}[htb]
\vspace*{-.2cm}
                 \insertplot{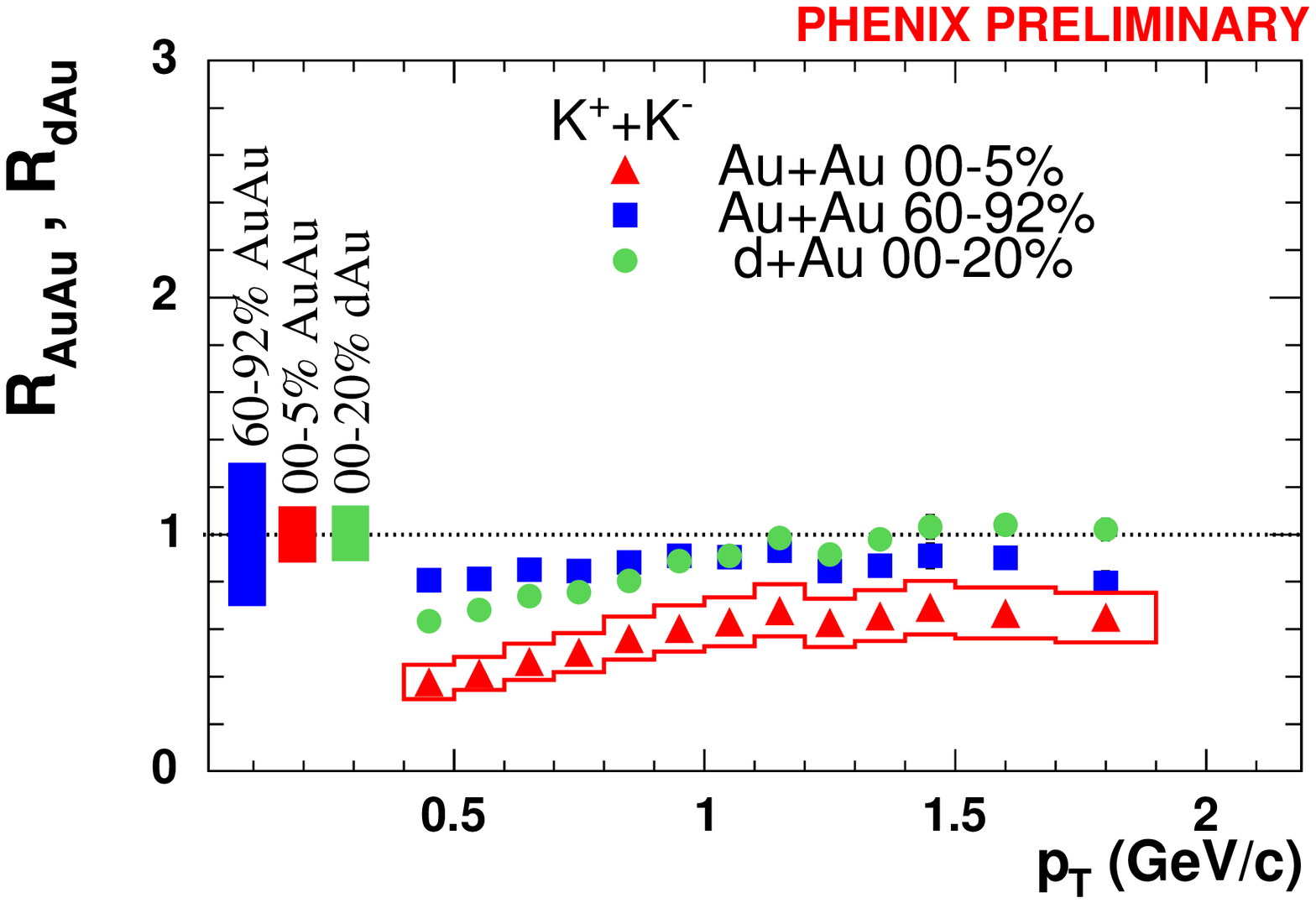}
\vspace*{-1cm}
\caption[]{ Nuclear modification factors for kaons,
comparing central and peripheral Au+Au collisions to central d+Au.
}
\label{fig:rauaudauk} 
\end{figure} 

Figures~\ref{fig:rauaudaupi}, \ref{fig:rauaudaup} and \ref{fig:rauaudauk}
 compare the nuclear modification factors for pions, kaons and 
(anti)protons in Au+Au and d+Au collisions. Central and
peripheral Au+Au collisions are compared to central d+Au collisions,
which have similar number of binary collisions as the peripheral
Au+Au sample. Pions show a much lower $R_{AA}$ at high $p_T$ in central
than in peripheral Au+Au collisions, as expected from the large
energy loss suffered by the quarks in central collisions.
The nuclear modification factor is 
slightly larger in d+Au than in peripheral Au+Au, despite the comparable
number of binary collisions, but we can not draw a definite conclusion due to 
the large systematic error of the peripheral $R_{AA}$.

The proton and antiproton nuclear modification factors
show a quite different trend, however. The Cronin effect, 
larger than 1.0 at higher $p_T$ values, is independent of 
centrality in Au+Au collisions. This feature was already 
observed as binary collision scaling
of proton and antiproton production in the central/peripheral
collision yield ratios~\cite{ppg015}. The Cronin effect in d+Au
is at least as large as in peripheral Au+Au. The difference
indicates that baryon production must involve a complex
interplay of processes in addition to initial state 
nucleon-nucleon collisions.

\section{Conclusions}\label{concl}

Results on identified particle production in
p+p, d+Au and Au+Au collisions at  $\sqrt{s_{NN}}$~=~200~GeV with the PHENIX spectrometer
are presented. The Cronin effect in d+Au collisions for charged pions is small, but non-zero.
The proton to pion ratio in d+Au is similar to that in peripheral Au+Au,
while the corresponding ratio in p+p is somewhat lower.
The nuclear modification factor in d+Au for protons shows a larger 
Cronin effect than that for pions but is not large enough
to account for the abundance of protons in central Au+Au collisions.
The difference between pions and protons does, however, indicate 
that the Cronin effect is not simply multiple scattering of the incoming partons.
$R_{AA}$ for protons and antiprotons confirms previous
observations that the production of high $p_T$ baryons in Au+Au
scales with the number of binary nucleon-nucleon collisions.

\vfill\eject

\begin{thebibliography}{99}

\def\Journal#1#2#3#4{{#1}{\bf #2}, #3 (#4)}
\def\IJMPA{{Int. J. Mod. Phys.}~{\bf A}}
\def\EPJ{{Eur. Phys. J.}~{\bf C}}
\def\JPG{{J. Phys}~{\bf G}}
\def\JHEP{{J. High Energy Phys.}~}
\def\NCA{Nuovo Cimento~}
\def\NIM{Nucl. Instrum. Methods~}
\def\NIMA{{Nucl. Instrum. Methods}~{\bf A}}
\def\NPA{{Nucl. Phys.}~{\bf A}}
\def\NPB{{Nucl. Phys.}~{\bf B}}
\def\PLB{{Phys. Lett.}~{\bf B}}
\def\PLC{Phys. Repts.\ }
\def\PRL{Phys. Rev. Lett.\ }
\def\PRL{Phys. Rev. Lett.\ }
\def\PRD{{Phys. Rev. D}}
\def\PRC{{Phys. Rev.}~{\bf C}}
\def\ZPC{{Z. Phys.}~{\bf C}}


\bibitem{ppg003} K.~Adcox {\it et al.}  [PHENIX Collaboration],
			\PRL{\bf 88}, 022301 (2002).

\bibitem{ppg014} S.S.~Adler {\it et al.} [PHENIX Collaboration], 
  \Journal{\PRL}{91}{072301}{2003}. 

\bibitem{ppg006} K.~Adcox {\it et al.}  [PHENIX Collaboration],
			\PRL{\bf 88}, 242301 (2002).

\bibitem{ppg015} S.S.~Adler {\it et al.} [PHENIX Collaboration], 
\PRL{\bf 91}, 172301 (2003).

\bibitem{ppg026} S.S. Adler {\it et al.} [PHENIX Collaboration].
			\PRC{\bf 69}, 034909 (2004).

\bibitem{Gyu90} M.~Gyulassy and M.~Pl\"umer, \PLB{\bf 243}, 432 (1990); 
X.N.~Wang and M.~Gyulassy, \PRL{\bf 68}, 1480 (1992).

\bibitem{BDMPS} R.~Baier, D.~Schiff and B.G.~Zakharov, Annu. Rev. Nucl. Part. Sci. {\bf 50}, 37 (2000), 
and references therein.

\bibitem{colorglas}
D. Kharzeev, E. Levin and L. McLerran, \PLB{\bf561}, 93 (2003).

\bibitem{cronin} J.W.~Cronin {\it et al.},  \Journal{\PRD}{11}{3105}{1975}.

\bibitem{straub} P.B.~Straub {\it et al.} \Journal{\PRL}{68}{452}{1992}.

\bibitem{antreasyan} D.~Antreasyan {\it et al.}, \Journal{\PRD}{19}{764}{1979}.

\bibitem{LevPetersson} M. Lev and B. Petersson, \ZPC{\bf 21}, 155 (1983).

\bibitem{Accardi} A. Accardi and M. Gyulassy, \PLB{\bf 586}, 244 (2004).

\bibitem{PappLevai} G.~Papp, P.~Levai, G.I.~Fai, \PRC{\bf 61}, 021902 (2000).

\bibitem{Vitev} I.~Vitev, M.~Gyulassy, \PRL{\bf89},  252301 (2002).  

\bibitem{Wang} X.N.~Wang, \PRC{\bf 61}, 064910 (2000).   

\bibitem{boris} B.Z. Kopeliovich, J. Nemchik, A. Schaefer and
A.V. Tarasov, \PRL{\bf 88} 232303 (2002).

\bibitem{Hwa} R.C. Hwa and C.B. Yang, \PRC{\bf 70}, 037901 (2004).

\bibitem{nim_phenix} K. Adcox {\it et al.}, \NIM{\bf A499} 469 (2003). 
		
\bibitem{glauber} R.J.~Glauber and G.~Matthiae, \NPB{\bf 21}, 135  (1970).

\bibitem{ppg036} S.S.~Adler {\it et al.} [PHENIX Collaboration], \Journal{\PRL}{94}{082302}{2005}. 

\bibitem{UA5} 
R.E.~Ansorge {\it et al.} [UA5 Collaboration], 
Nucl. Phys. B{\bf 328}, 36 (1989).

\bibitem{heinzpaper}
J.~Adams  {\it et al.}, [STAR Collaboration], nucl-ex/0403020, (2004).

\bibitem{heinzppt}
M.~Heinz {\it et al.}, [STAR Collaboration],
J.\ Phys.\ G {\bf 31}, S141 (2005)

\bibitem{cai}
X.~Cai [STAR Collaboration],
8th International Conference on Strangeness in Quark Matter, 
Cape Town, South Africa (2004).

\bibitem{mtscaling}
R.~Witt [STAR Collaboration], nucl-ex/0403021 (2004).

\bibitem{ppg028} S.S. Adler {\it et al.} [PHENIX Collaboration],
			\PRL{\bf 91}, 072303 (2003).

\bibitem{starpiddau} J. Adams {\it et al.} [STAR Collaboration],
		nucl-ex/0309012 (2003).


\end{thebibliography}
\end{document}